%% file: prd_v50.tex
\documentstyle[prd,aps,epsfig,twocolumn,amstex,floats]{revtex}

\input d0_style

\newcommand{\GEAN}{{\sc geant}}

\def\wb{$W$}
\def\zb{$Z$}
\def\ee{$e^+e^-$}
\def\tev{TeV}
\def\Dzero{D\O}

\begin{document}
\onecolumn

\title{ A direct measurement of the W boson decay width }

\include{list_of_authors_r1}

\maketitle

\begin{abstract}
Based on 85 pb$^{-1}$ data of \ppbar\ collisions at $\sqrt{s}=1.8$ \tev\  
collected using the D\O\ detector at Fermilab during the 1994-1995 run of 
the Tevatron, we present a direct measurement of the total decay width of 
the \wb\ boson, $\Gamma_W$.
The width is determined from the transverse mass spectrum in the 
$W \rightarrow e+\nu_e$ decay channel and found to be 
$\Gamma_W = 2.23^{+ 0.15}_{- 0.14}$(stat.)$\pm 0.10$(syst.)GeV,
consistent with the expectation from the standard model.
\end{abstract}

\vskip 0.5in
\leftline{March 30, 2002}

\twocolumn

\clearpage

\section{Introduction}
\label{sec-introduction}
The theory that describes the fundamental particle interactions is called 
the standard model (SM). The standard model is a gauge field theory
that comprises the Glashow-Weinberg-Salam (GWS) 
model~\cite{SM_Glashow,SM_Weinberg,SM_Salam} of the weak and electromagnetic 
interactions and quantum chromodynamics (QCD)~\cite{QCD_Bardeen,QCD_Gross,QCD_Weinberg}, 
the theory of the strong interactions. 
The discovery of the \wb\ ~\cite{UA1_W_discovery,UA2_W_discovery} and 
\zb\ ~\cite{UA1_Z_discovery,UA2_Z_discovery} bosons in 1983 by the UA1 
and UA2 collaborations at the CERN \ppbar\ collider provided a direct 
confirmation of the unification of the weak and electromagnetic interactions.  
Experiments have been 
refining the measurements of the characteristics of the \wb\ and \zb\ bosons.
The total decay width of $W$ boson, $\Gamma_W$, is given 
in the SM in terms of the masses of the gauge bosons 
and their couplings to their decay products. 

In \ppbar\ collisions, \wb\ bosons are produced by processes of the type 
$u\overline{d}$ or $\overline{u}d \rightarrow W$, followed by 
subsequent leptonic or hadronic decay:
$W \rightarrow \ell \nu$ or
$W \rightarrow q^\prime \overline{q}$,
\noindent where $\ell = e,\mu,\tau $, and $q^\prime$ or $q$ represent one of the quarks 
$u$, $d$, $c$, $s$ or $b$ (but not $t$ since top quark is heavier than the \wb\ boson). 

At lowest order in perturbation theory, the SM predicts the partial decay 
width $\Gamma(W \rightarrow e\nu)$ of $W \rightarrow e\nu$ to be 
$\Gamma(W \rightarrow e\nu) = g^2M_W/48\pi$~\cite{epj:groom}.
Including radiative corrections, this can be rewritten as:
\begin{equation}
\Gamma(W \rightarrow e\nu) = {{G_FM_W^3\over 6\sqrt{2}\pi} (1+\delta_{\rm SM})}.
\end{equation}

\noindent where $G_F/\sqrt{2}=g^2/8M_W^2$, $g$ is the charged current 
coupling, and $M_W$ is the mass of the $W$ boson. The SM radiative correction, 
$\delta_{\rm SM}$, is calculated~\cite{prd:rosner} to be less than 
$\frac{1}{2}\%$. By using the experimental values of $G_F$ (measured 
from muon decay~\cite{fconst}) and $M_W$ (measured at the Fermilab Tevatron 
collider~\cite{cdf:mu,wmassprd} and LEP2~\cite{wmassl3,wmassalph,wmassopal,wmassdlph}), 
the predicted partial width 
is~\cite{epj:groom} $\Gamma(W \rightarrow e\nu) = 226.5 \pm 0.3\hspace{0.2cm} \text{MeV}$

A $W$ boson has three leptonic decay channels and two dominant hadronic 
decay channels, $W\rightarrow e\bar{\nu}, \mu\bar{\nu}, \tau\bar{\nu}$
and $qq'$, where q is $u$ or $c$, and $q'$ is the appropriate CKM mixture 
of $d$ and $s$. Other hadronic decay channels are greatly 
suppressed by CKM off-diagonal matrix elements. Considering the three 
color charges for quarks, these nine leptonic and hadronic channels 
yield a total width of $\approx 9\Gamma(W \rightarrow e\nu)$.
Including QCD corrections, the leptonic decay branching ratio is 
$B(W\rightarrow e\nu) = 1/\{3+6[1+\alpha_s(M_W)/\pi+{\cal O}(\alpha_s^2)]\}$, 
leading to the SM prediction for the full width of the $W$ boson~\cite{epj:groom}
of
$\Gamma_W = 2.0921 \pm 0.0025 \hspace{0.2cm} \text{GeV}$.

Historically, the accurate determination of the width of the $W$ boson was 
available through
an indirect measurement using the ratio $\cal R$ of the $W\rightarrow e\nu$ and 
$Z\rightarrow ee$ cross sections:

\begin{eqnarray}
\cal R & =& {\sigma(p\bar{p} \rightarrow W+ X)\cdot Br(W \rightarrow e\nu) \over
            \sigma(p\bar{p} \rightarrow Z + X)\cdot Br(Z \rightarrow ee)}\cr
       & =& {\sigma_W \over \sigma_Z}\cdot{{Br(W\rightarrow e\nu)}\over{Br(Z\rightarrow ee)}}.
\end{eqnarray}
A measurement of $\cal R$, together with a calculation~\cite{mrsd:plb} of the 
ratio of production cross sections $\sigma_W/\sigma_Z$ and the measurement of 
the branching faction $Br(Z\rightarrow ee) = \Gamma (Z\rightarrow ee)/\Gamma (Z)$ 
from the CERN $e^{+}e^{-}$ collider (LEP)~\cite{cern:lep}, can be used to 
extract the $W$ boson leptonic branching ratio 
$Br(W\rightarrow e\nu) = \Gamma (W \rightarrow e\nu)/\Gamma(W)$, which, 
in turn, yields the full width of the $W$ boson from calculated partial decay 
width $\Gamma (W \rightarrow e\nu)$. Thus, in this indirect measurement, 
calculations of $\sigma_W/\sigma_Z$ and the partial width 
$\Gamma (W \rightarrow e\nu)$ yield $\Gamma_W$ in the context of the SM.
This method was first used by the UA1~\cite{pl:ua1} and UA2~\cite{pl:ua2}
collaborations. More recently, the CDF~\cite{cdf:ratio} and \D0~\cite{wzcsprd} 
collaborations obtained $\Gamma_W = 2.064\pm 0.084$~\gev~and 
$\Gamma_W = 2.169\pm 0.079$~\gev, respectively, using this technique.

\begin{figure}[htpb!]
\vspace{0.5in}
\centerline{\psfig{figure=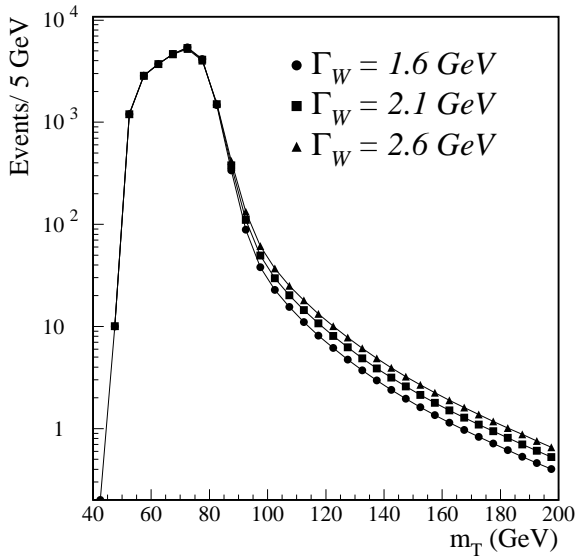,width=3.0in}}
\vspace{-0.02in}
\caption{Monte Carlo simulations of the transverse mass spectrum for 
different $W$ boson widths. The selections, $E_T(e) >$ 25 GeV and 
$E_T(\nu) >$ 25 GeV, are applied to MC sample. The circles show the spectrum for 
$\Gamma_W = 1.60$~\gev, the squares for $\Gamma_W = 2.10$ GeV, 
and triangles for $\Gamma_W = 2.60$~\gev. Distribution are 
normalized arbitrarily in the transverse mass region shown.}
\label{fig:difwidth}
\end{figure}

The value of $\Gamma_W$ can also be obtained from the line shape of 
the transverse mass $m_T$ of the $W$ boson, because the Breit-Wigner 
(width) component of the line shape falls off more slowly at high 
$m_T$ than the resolution component does~\cite{prd:rosner}. 
The transverse mass is given by 
\begin{equation}
m_T = \sqrt{2E_T^eE_T^\nu[1-\cos(\phi^e-\phi^\nu)]}
\end{equation}
where $E_T^e$ and $E_T^\nu$ are the transverse energies,
and $\phi^e$ and $\phi^\nu$ are the azimuthal angles, of the 
electron and neutrino, respectively. The transverse mass has a kinematic
upper limit at the value of $M_W$, and the shape of the 
$m_T$ distribution at this upper limit, called the ``Jacobian edge,''
is sensitive to $\Gamma_W$~\cite{vdb:rjnp}. Using this technique,  
the CDF collaboration reported~\cite{cdf:wwidth} a measurement of 
$\Gamma_W = 2.05\pm 0.10$(stat.)$\pm$0.08(syst.)~\gev. 
Figure~\ref{fig:difwidth} shows the $m_T$ spectrum shape expected for 
different values of $\Gamma_W$ and indicates the sensitivity of 
the tail of the transverse mass distribution to $\Gamma_W$.
Clearly, the effect is greatest in the region above $m_W$.

The direct measurement of $\Gamma_W$ complements the indirect measurement 
through $\cal R$ in several ways:
\begin{itemize}
  \item Theoretical inputs for $\sigma_W/\sigma_Z$ and $\Gamma(W\rightarrow e\nu)$, 
        which may be sensitive to non-SM coupling of the $W$ boson, are not needed.
  \item The direct measurement explores the region above the $W$ boson mass pole, where 
        possible new phenomena such as an additional heavy vector boson ($W'$) can 
        contribute.
  \item It is desirable to have more than one method of measuring 
        a given property. The sources of systematic errors in the two methods 
        are different, and the direct method will be important when the measurement 
        through $\cal R$ becomes limited by systematic uncertainty.
\end{itemize}

The paper is organized as follows. In Sec.\ II, we give a brief description 
of the \D0\ detector. Particle identification and event selection are
discussed in Sec.\ III. The analysis procedure, including background estimation 
and Monte Carlo simulation, is 
described in Sec.\ IV, and the conclusions are presented in Sec. V. 
For more detailed information on this analysis, see Ref.~\cite{xuthesis}.

\section {The \Dzero\ Detector}
\label{sec-exper}
\subsection{Experimental Apparatus}
The \Dzero\ detector~\cite{d0nim} comprises three major systems. 
The innermost of these is a non-magnetic
tracker used in the reconstruction of charged particle tracks. The tracker 
is surrounded by  central and forward uranium/liquid-argon sampling 
calorimeters. These calorimeters are used to identify electrons, photons, 
and hadronic jets, and to reconstruct their energies. The calorimeters are 
surrounded by a muon spectrometer used in the identification of muons and 
the reconstruction of their momenta. We use a coordinate system 
($\rho, \theta, \phi$) where $\rho$ is the perpendicular distance from 
the beam line, $\theta$ is the polar angle measured relative to the proton 
beam direction $z$, and $\phi$ is the azimuthal angle. The pseudorapidity 
$\eta$ is defined as $-\ln(\tan\frac{\theta}{2})$. For this analysis, 
the relevant components are the tracking system and the calorimeters.  

The central tracking system provides a measurement of the energy loss 
due to ionization ($dE/dx$) for tracks within its tracking volume.  
This information is used to help distinguish prompt electrons from 
\ee\ pairs due to photon conversions. 

The structure of the calorimeter has been optimized to distinguish 
electrons and photons from hadrons and to measure their energies.
It is composed of three sections: the central calorimeter (CC), and 
two end calorimeters (EC). The $\eta$-coverage for electrons used 
in this analysis is $\mid \eta\mid<1.1$~\cite{etadef} in the CC region, 
which consists of 32 $\phi$ modules. The calorimeter is segmented 
longitudinally into three sections, the electromagnetic 
calorimeter (EM), the fine hadronic calorimeter (FH), and the 
coarse hadronic calorimeter (CH). The EM calorimeter is subdivided 
longitudinally into four layers (EM1--EM4). The first, second and
fourth layers of the EM calorimeter are transversely divided into  
cells of size $\Delta\eta\times \Delta\phi = 0.1 \times 0.1 $.
The electromagnetic shower maximum occurs in the third layer, which 
is divided into finer units of $ 0.05 \times 0.05 $ to improve 
the measurement of the shower shape and spatial resolution. There 
are 16 FH modules and 16 CH modules in $\phi$. The fine hadronic 
calorimeter is subdivided longitudinally into three fine hadronic 
layers (FH1--FH3), and there is only one coarse hadronic layer.  
\subsection{Trigger}
\label{sec:trig}
The \Dzero\ trigger has three levels, each applying
increasingly more sophisticated selection criteria to an event. 
The lowest level trigger, Level 0, uses scintillation counters 
located on the inner faces of the forward calorimeters to signal 
the presence of an inelastic \ppbar\ collision.
Data from the Level 0 counters, the calorimeter, and the muon chambers 
are sent to the Level 1 trigger, which provides a trigger 
on total transverse energy (\et), missing transverse energy (\met), \et\ of 
individual calorimeter towers, and/or the presence of a muon. These 
triggers operate in less than  $3.5\;\mu\mbox{s}$, the time between bunch 
crossings. Some calorimeter and muon-based triggers require additional time, 
which is provided by a Level 1.5 trigger system.

Level 1 (and 1.5) triggers initiate a Level 2 trigger system that consists 
of a farm of microprocessors. These microprocessors run simplified versions 
of the off-line event reconstruction algorithms to select events of interest.

\section {Particle Identification and Event Selection}

\label{partid}
This analysis relies on the \Dzero\ detector's ability to identify 
electrons and to associate the undetected energy with neutrinos. We 
use both $W \rightarrow e\nu$ and $Z\rightarrow e^{+}e^{-}$ candidate 
samples for this analysis. The $W$ boson candidate sample provides 
the signal events, while the $Z\rightarrow e^{+}e^{-}$ candidate 
sample is used to calibrate both the data and the Monte Carlo (MC) 
simulation. Candidate \wb\ and \zb\ events are identified by the 
presence of an electron and a neutrino, or by the presence of two 
electrons with an invariant mass consistent with the mass of the \zb\ 
boson, respectively. Electrons from \wb\ and \zb\ boson decays 
typically have large transverse energy and are isolated from other 
particles. They are associated with a track in the tracking system 
and with a large deposit of energy in one of the EM calorimeters. 
Neutrinos do not interact in the detector, and thus create an 
apparent transverse energy imbalance in an event. For each \wb\ 
boson candidate event, we measure the energy imbalance in the plane 
transverse to the beam direction (\met), and attribute this to the 
neutrino. The following sections provide a brief summary of the 
procedure~\cite{wzcsprd} used in this analysis.

\subsection{Electron Identification} 
Identification of electrons starts at the trigger level with the selection 
of clusters of electromagnetic energy. At Level 1, the trigger searches for 
EM calorimeter towers ($\Delta\phi\times\Delta\eta=0.2\times0.2$) with signals that 
exceed predefined thresholds. $W$ boson triggers require that the energy 
deposited in a single EM calorimeter tower exceed 10 {GeV}. Those events 
that satisfy the Level 1 trigger are processed by the Level 2 filter. The 
trigger towers are combined with the energy in the surrounding calorimeter 
cells within a window of $\Delta\phi\times\Delta\eta=0.6\times0.6$. 

Events are selected at Level 2 if the transverse energy in this window 
exceeds 20 GeV. In addition to the $E_T$ requirement, the longitudinal 
and transverse shower shapes are required to match those expected for 
electromagnetic showers. The longitudinal shower shape is described by 
the fraction of the energy deposited in each of the four EM layers of 
the calorimeter. The transverse shower shape is characterized by energy 
deposition patterns in the third EM layer. The difference between the 
energies in concentric regions covering $0.25\times 0.25$ and $0.15\times 0.15$ 
in $\Delta\eta\times\Delta\phi$ must be consistent with that expected 
for an electron~\cite{d0nim}. 

In addition, the electron candidates are required to deposit at least 90\% 
of their energy in the EM section of the calorimeter and to be isolated 
from other calorimetric energy deposits. To be considered isolated, 
electrons must satisfy the isolation requirement, 
$f_{\text{iso}}<0.15$, where $f_{\text{iso}}$ is defined as:

\begin{equation}
f_{\text{iso}}=\frac{E_{{\rm total}}(0.4)-E_{{\rm EM}}(0.2)}{E_{{\rm EM}}(0.2)}
\end{equation}
in which $ E_{{\rm total}}(0.4)$ is the total energy, and $E_{{\rm EM}}(0.2)$ 
the electromagnetic energy, in cones of radius 
$R=\sqrt{(\Delta\eta)^2+(\Delta\phi)^2} =0.4$ and $0.2$, respectively.
This enhances the signal expected from isolated electrons in $W$ and $Z$ 
boson decay.

Having selected events with isolated electromagnetic showers at the 
trigger level, we first define ``loose''  electron for the purpose 
to study the background. Those EM clusters are require to locate within 
the center 80\% of a calorimeter module, have an associated track 
in the central tracking volume and $|\eta|<1.1$. To avoid areas of 
reduced response between neighboring calorimeter modules, the azimuthal 
angle of electrons is required to be at least $\Delta \phi = 0.10 \times 2\pi/32$ 
radians away from the position of a module boundary.
We further impose a set of off-line tighter criteria to identify
electrons, thereby reducing the background from QCD multijet events. 
The first step in identifying an electron is to form a cluster around 
the trigger tower using a nearest neighbor algorithm. As at the trigger 
level, the cluster is required to be isolated ($f_{\text{iso}}<0.15$). To 
increase the likelihood that the cluster is due to an electron and 
not a photon, a charged track from the central tracking system is required 
to point to the center of the EM cluster.
We extrapolate the track to the third EM layer of the calorimeter and 
calculate the distance between the extrapolated track and the cluster 
centroid along the azimuthal direction ($\rho \Delta \phi$) and in the 
$z$-direction ($\Delta z$). The position of cluster centroid is defined 
at the radius of the third EM layer of the calorimeter.
The $z$ position of the event vertex is then defined by the line connecting 
the centroid position of the EM cluster to associated one in in the 
central tracking system and 
extrapolated to the beam line. The electron \et\ is calculated using 
this vertex definition~\cite{wzcsprd}.
The variable 
\begin{equation}
\sigma_{\text{trk}}^2=\left( \frac{\rho\Delta \phi}{\sigma_{\rho \phi}}\right)^2 +
\left( \frac{\Delta z}{\sigma_z}\right)^2. \label{eq:trk_pos}
\end{equation}
where  $\sigma_{\rho \phi}$ and $\sigma_z$ are the respective track 
resolutions, quantifies the quality of the match. A requirement of 
$\sigma_{\text{trk}} < 5$ is imposed on the data.
These clusters are then subjected to a 4-variable likelihood 
test~\cite{topmassprd,miguel}. The four variables are:
\begin{itemize}
  \item A $\chi^2$ comparison of the shower shape with the expected shape 
        of an electromagnetic shower, computed using a $41$-variable 
        covariance matrix~\cite{covprd} for the energy depositions in the 
        cells of the electromagnetic calorimeter and the location of event vertex.

  \item The electromagnetic energy fraction, defined as the 
        ratio of shower energy in the EM section of the calorimeter 
        relative to the sum of EM energy plus the energy in the 
        first hadronic section of the calorimeter.

  \item A comparison of the track position to the position of cluster centroid, as
        defined in Eq.~\ref{eq:trk_pos}.

  \item The ionization, $dE/dx$, along the track. This is used to reduce  
	contamination due to $e^+e^-$ pairs from photon conversions,
        mainly from jets fragmenting into neutral pions. The $e^+e^-$ pair 
        from photon conversion has a double value of $dE/dx$ for a geniue electron
        due to two overlapping tracks,
\end{itemize}
To good approximation, these four variables are independent of each other 
for electron showers. 
Electrons that satisfy all above criteria are called 
``tight'' electrons. 

Electron energies are corrected for the underlying event energy that enter
into the electron windows. The electromagnetic energy 
scale is determined in the test beam data, and adjusted to make the peak of 
the $Z\rightarrow e^{+}e^{-}$ invariant mass agree with the known 
mass of the \zb\ boson~\cite{cern:lep}. We found it to be $0.9545 \pm 0.0008$. 
The electron energy scale is discussed in detail in Ref.~\cite{wmassprd}.
\subsection{Missing Transverse Energy} 
The primary sources of missing energy in an event include the neutrinos 
that pass through the calorimeter undetected and the 
calorimeter resolution. The energy imbalance is measured 
only in the transverse plane because of the lost particles emitted at small angles
(within the beam pipes). 
The missing transverse energy is calculated by taking the negative of the 
vector sum of the transverse energy in all of the calorimeter cells. 
This gives both the magnitude and direction of \met, allowing the 
calculation of the transverse mass of the $W$ boson candidates.

\subsection{Event Selection}
\begin{figure}[htpb!]
\centerline{\psfig{figure=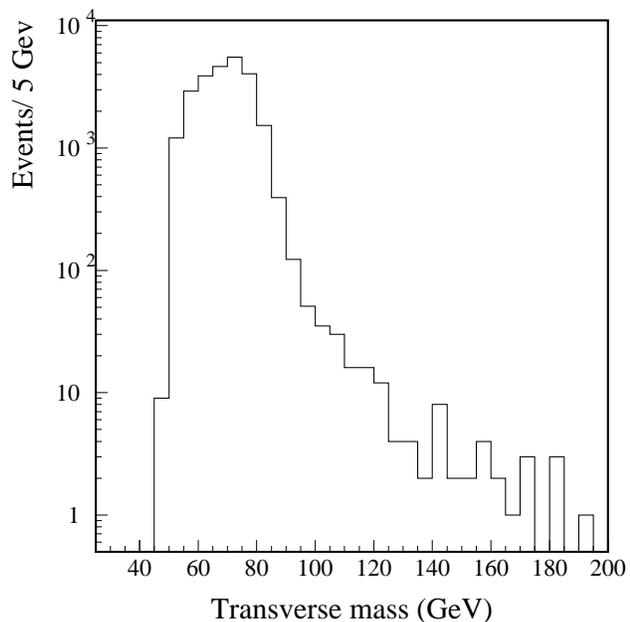,width=3.3in}}
\vspace{-0.02in}
\caption{Transverse mass distribution of $W \rightarrow e\nu$ event candidates.
}
\label{fig:wmt}
\end{figure}

The $W$ boson data sample used in this analysis was collected during 
the 1994--1995 run of the Fermilab Tevatron collider, and  
corresponds to an integrated luminosity of $85.0\pm 3.6\;\rm pb^{-1}$.
Events are selected by requiring one tight electron in the central 
calorimeter $(\mid \eta \mid <1.1)$~\cite{etadef} with \et\ $>25$ GeV.
In addition, events are required to have \met\ $>25$ GeV and $W$ 
transverse momentum $p_T(W) < 15$ GeV,
which is combined transverse momentum of electron and \met\ (neutrino). 
After applying all of the described selections, a 
total of 24487 $W$ boson candidates is selected. There are 24479 candidates 
in the region 0 - 200 GeV, while 8(2) candidates have $m_T >$200(250) GeV. 
Figure~\ref{fig:wmt} shows the transverse mass distribution of the 
$W \rightarrow e\nu$ candidates.

\begin{figure}[htpb!]
\vspace{0.5in}
\centerline{\psfig{figure=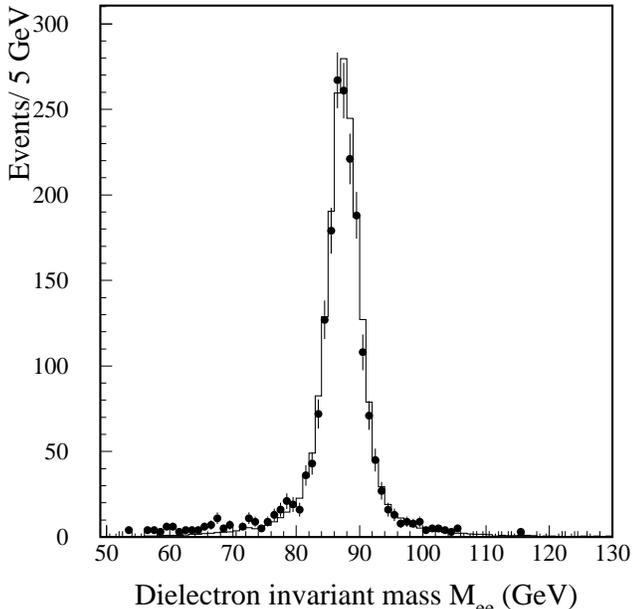,width=3.3in}}
\vspace{-0.02in}
\caption{Invariant mass distribution of $Z\rightarrow e^{+}e^{-}$ events 
compared to Monte Carlo simulation. The histrogram is the MC and the black
dot with error bar is the data. The $Z\rightarrow e^{+}e^{-}$ candidate 
require both electrons be in the CC.}
\label{fig:zee}
\end{figure}

Candidates for the process $Z\rightarrow e^{+}e^{-}$ are required 
to have two tight electrons, each with \et\ $>$ 25 GeV in the CC. The invariant mass 
of the dielectron pair is required to satisfy 
60 GeV $< m_{ee} <$ 120 GeV. A total of 1997 \zb\ boson 
candidates is selected. Figure~\ref{fig:zee} shows the 
invariant mass distribution of the $Z\rightarrow e^{+}e^{-}$ candidates.
\section{Analysis Procedure}
\label{Analysis}

In this section, we describe the Monte Carlo simulation program used to model 
the transverse mass spectrum. The background from the dominant 
processes that can mimic the $W \rightarrow e\nu$ signal is also estimated.
We compare the data with the expectation from the Monte Carlo simulation and 
extract the decay width of the $W$ boson using log-likelihood fits to the $W$ boson 
transverse mass distribution.

\subsection {Monte Carlo Simulation \label{sec:mc}}
The transverse mass spectrum for the $W$ boson is modeled in three steps: 
$W$ boson production, $W$ boson decay, and a parameterized 
detector simulation~\cite{wmassprd,flattum,wmass:thesis}. 

We first simulate the production of the $W$ boson by generating 
its four momentum and other event characteristics, such as the 
$z$-position of the interaction vertex and the run luminosity. The
luminosity is used to parameterize luminosity-dependent effects. 
To lowest-order, the mass of the $W$ boson follows the
Breit-Wigner distribution:
\begin{equation}
\sigma(Q) = {\cal L}_{q\bar{q}}(Q) \frac{Q^2}{(Q^2-M_W^2)^2 + Q^4\Gamma_W^2/M_W^2}. \label{eq_bw}
\end{equation}
where $Q$ is the invariant mass of $W$ boson, $M_W$ is the pole mass and $\Gamma_W$ the 
decay width of the $W$ boson, and ${\cal L}_{q\bar{q}}(Q)$ is called the parton luminosity. 
To evaluate ${\cal L}_{q\bar{q}}(Q)$, we generate $W \rightarrow e\nu$ events using the 
leading-order {\sc resbos}~\cite{resb} event generator and the different PDF models 
described in Refs.~\cite{ly,ARtheory}, and then select the events using the same kinematic 
and fiducial constrains as for the $W$ and $Z$ boson data samples. The resulting event 
distribution is proportional to the parton luminosity, which we parameterize with the 
function~\cite{parton}:
\begin{equation}
{\cal L}_{q\bar{q}}(Q) = \frac{e^{-\beta Q}}{Q}.
\end{equation}
where $\beta$ is obtained from a fit of the MC events to Eq.~\ref{eq_bw}.

The decay of the $W$ boson is simulated in the MC and used to calculate the transverse 
momentum of the electron and other decay products. Any radiation from the
decay electron or from the $W$ boson can bias the measurement and has to be taken
into account. $W \rightarrow \tau\nu \rightarrow e\nu\bar{\nu}\bar{\nu}$ events are 
indistinguishable from $W \rightarrow e\nu$ and are also included in the 
model, using a branching ratio of 
$Br(\tau \rightarrow e\nu\bar{\nu})/[1+Br(\tau \rightarrow e\nu\bar{\nu})]=0.151$. 

Finally, we apply a parameterized detector simulation to the momenta
of all decay products to simulate any observed recoil jets and 
electron momenta. The parameters giving the electron and recoil 
system response of the detector are fixed using data, which include 
$Z$ bosons and their 
recoil jets, to study calorimeter response and resolution. The response 
to jets and electrons is parameterized as a function of energy and angle. Also 
included in the detector parameterization are effects due to the 
longitudinal spread of the interaction vertex and the 
luminosity-dependent response of the detector caused by multiple collisions.

Uncertainties in the input parameters to the MC will eventually limit the 
accuracy of the width measurement of the $W$ boson. To study the 
uncertainties, we allow these input parameters to vary by one standard 
deviation and re-generate the corresponding transverse mass spectrum. We then 
fit it with a nominal MC template. If the positive and negative variations 
of the width of the $W$ boson with respect to a parameter are not symmetric, 
the larger value is used for the uncertainty. This estimation is used to
estimate the impact of the electron energy resolution, hadronic energy resolution, 
electron energy scale, hadronic energy scale, dependence on the $W$ boson mass, electron
angular calibration,  and radiative corrections. Detailed studies of these 
parameters can be found in Ref.~\cite{wmassprd}. The uncertainties on 
$\Gamma_W$ from the electron energy resolution and scale are 27 MeV and
41 MeV, respectively. The uncertainties from the hadronic energy resolution
and scale lead to variations in $\Gamma_W$ of 55 MeV and 22 MeV, 
respectively. The error on the $W$ boson mass of 37 MeV has an effect of 15 MeV on
$\Gamma_W$. The uncertainties from radiative decay and electron 
angular calibration correspond to 10 MeV and 9 MeV, respectively.

Uncertainties on $\Gamma_W$ also arise from uncertainties in the production 
model and the parton distribution functions (PDFs). The uncertainty from the 
former is determined from the upper and lower limits~\cite{ly} of the most 
uncertain parameter in the model. This leads to an uncertainty of 28 MeV due 
to parton luminosity and 12 MeV due to uncertainty in the transverse momentum 
of the $W$ boson in the model. There are several PDF models currently in use.
The uncertainty due to variation in PDFs is determined by using different PDFs, 
including MRSA~\cite{MRSA}, CTEQ4M and CTEQ5M~\cite{CTEQ}, and finding the 
largest excursion from the value of $\Gamma_W$ determined using the MRST PDF 
set~\cite{MRST}, leading to a variation of 27 MeV. The value quoted for $\Gamma_W$ 
is determined using the MRST PDFs. We chose MRST so that the results can be 
consistent with \D0 mass analysis~\cite{wmassprd}.

\subsection {Backgrounds\label{sec:bg}}

Backgrounds to $W\rightarrow e\nu$ can affect the shape of the $m_T$ 
spectrum and skew the measurement of $\Gamma_W$. We account for this 
by estimating the background as a function of $m_T$ and adding this to 
the $m_T$ distribution of the $W$ boson from the Monte Carlo.
The three dominant background sources are multijet events, $Z \to ee$, 
and $W \to \tau \nu$ decay products. The following 
describes how the backgrounds are estimated~\cite{xuthesis}. 

A large potential source of background is due to multijet events in 
which one jet is misidentified as an electron and the energy in the 
event is mismeasured, thereby yielding large \met. This background is 
estimated using jet events from data, following the procedure called the 
``matrix method,'' described in Ref.~\cite{wzcsprd,xuthesis,miguel}. The 
method uses two sets of data, each containing both signal and background. 
The first data set corresponds to the $W$ data sample in this analysis. 
The second set contains a different mix of signal and background which 
is obtained with loose electron criteria (described in Sec.\ III A). We 
summarize below the essence of this method used to estimate the multijet 
background.

\begin{figure}[!htbp]
\centerline{\psfig{figure=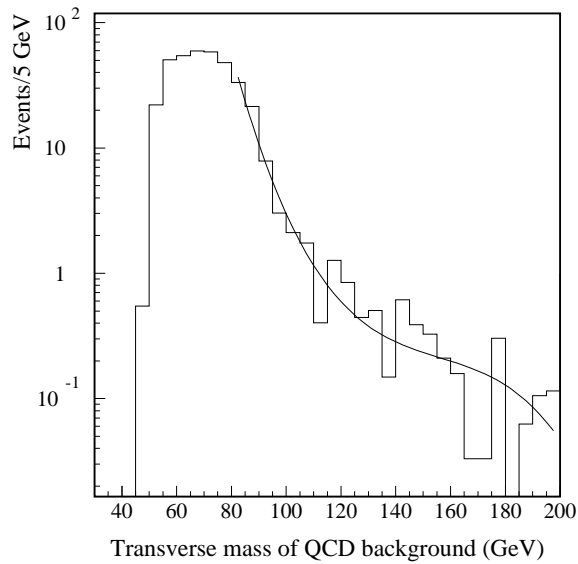, width=3.0in}}
\caption{ The transverse mass distribution for the multijet background.  
The line represents the results of the fit described in the text.}
\label{fig:bkgfit}
\end{figure}

The number of multijet background ($N^W_{BG}$) events in the tight electron 
$W$ data sample is given by 
\begin{equation}
N^W_{BG}=\epsilon_j\frac{\epsilon_s N_l-N_t}{\epsilon_s-\epsilon_j}.
\end{equation}
where $N_l$ and $N_t$ are the number of events in the $W$ boson samples 
satisfying loose and tight electron criteria, respectively. The tight 
electron efficiency, $\epsilon_s$, is the fraction of loose electrons 
that pass tight electron criteria, as determined by the $Z$ boson sample, where one 
electron is required to pass the tight selection criteria and the other 
serves as an unbiased probe for determining relative efficiencies. 
The electron efficiency is obtained to be $\epsilon_s = (86.3 \pm 1.2)\%$.
The jet efficiency $\epsilon_j$ is the fraction of loose ``electrons'' found 
in multijet events that also pass tight electron criteria. This sample 
is required to have $\met \le 15$ GeV to minimize the number of $W$ bosons 
contained in it. The results is $\epsilon_j = (5.83 \pm 0.25) \%$.
Once $\epsilon_s$ and $\epsilon_j$ are 
determined, we can extract the background-event distribution.
The ``electron'' and ``neutrino'' transverse momenta and energies are used to 
form the transverse mass, and this distribution is
shown in Figure~\ref{fig:bkgfit}. The total multijet background is estimated 
to be $368\pm32$ events in the region $m_T < 200$ GeV, 
with $25.4\pm2.2$ events in the range 90 GeV $< m_T < 200$ GeV. 

The background sample is smoothed in the region 85 
GeV $< m_T < 200$ GeV. We fit the distribution to 
an exponential function of the form
$f_{BG}$ = exp$(a_0 + a_1 x + a_2 x^2 + a_3 x^3)$.
The fitting parameters $a_0, a_1, a_2$ and $a_3$~\cite{bkgfitpara}
are used to generate the background 
distribution for the fit to the signal. For bins outside the 
fitted region, we use the original data itself, as shown in 
Figure~\ref{fig:bkgfit}.

%
%
Another source of background is due to $Z\rightarrow ee$ events
in which one electron is undetected.
This results in a momentum imbalance, with the event now being topologically
indistinguishable from $W \rightarrow e\nu$ events. This 
background is also estimated using Monte Carlo events.
The number of such $Z$ boson events present in the $W$ boson 
sample is calculated by applying the $W$ boson selection criteria 
to MC $Z \rightarrow ee$ events generated using {\sc herwig}~\cite{herwig} 
and processed through a {\sc geant}~\cite{geant} 
based simulation of the \D0 detector, and then overlaid with 
events from random $p$\pbar\ crossings. This is done to simulate the effect
of the luminosity on the underlying event.
Out of a total of 8870 $Z\rightarrow ee$ events, 48 pass the $W$ boson event 
selection. Normalizing the Monte Carlo sample to the size of the data 
sample for equivalent luminosity, we estimate 
that there are 102 $Z\rightarrow ee$ events in the data sample.

$W\rightarrow \tau \nu$ events in which the $\tau$ decays
into an electron and two neutrinos are indistinguishable from
$ W\rightarrow e\nu$ events on an event-by-event basis. 
Because $\tau$ undergoes a three-body decay, 
leading to a softer electron relative to $W \rightarrow e\nu$ events,
the acceptance is reduced greatly by the standard \et\ selection criteria. The size 
of this background is small, and it tends to add events with low values of $m_T$.
This background is determined using the $W\rightarrow e\nu$ 
Monte Carlo, modified to include the decay of the $\tau$ lepton. 
The events are then passed through the same detector simulation
used to model the $W\rightarrow e\nu$ signal. 

The shape and total amount of background affect the fit used to determine 
the width of $W$ boson. To estimate the uncertainty in $\Gamma_W$ due to the 
uncertainty in absolute background, we scale up (and down) the fitted number of 
background events by an amount that corresponds to the total uncertainty in the 
background. This gives an uncertainty of 15 MeV for $\Gamma_W$ extracted from 
the region 90 GeV $< m_T <$ 200 GeV.
To estimate the uncertainty in $\Gamma_W$ from the uncertainty in the shape of the background 
spectrum, we perform an ensemble study in which background is generated 
using a multinomial distribution. The multinomial distribution is defined by:
\begin{equation}
P(N_1,N_2,\ldots,N_{ch}) = N_{\text{total}}\prod_{i=1}^{ch}{{p_i^{N_i}}\over{N_i!}}.
\end{equation}
\noindent where $N_{\text{total}}$ is the total number of background events, 
$ch$ is the number of the bins, 
$p_i$ is the original distribution, and $N_i$ is numbers of events in $i$-$th$ bin.
The total background $N_{\text{total}}$ is kept at its central 
value, while the number of background events in each bin is allowed to fluctuate. 
The $W$ boson width is then recalculated with the new background distribution. 
The variation in $\Gamma_W$ is taken as the 
uncertainty. We found that this is 39 MeV for the fitted region of $m_T$.

\subsection{Likelihood Fitting}

We generate a set of Monte Carlo $m_T$ templates 
with $\Gamma_W$ varying from 1.55 GeV to 2.75 GeV at intervals of 50
MeV. These templates are normalized to the number of events in the
region of $m_T < 200$ GeV. The background distributions of multijet
and $Z \rightarrow ee$ events are added to
the templates and a binned likelihood is calculated for
data. The $m_T$ bin size is 5 GeV. The fitting region is chosen to be 
90 GeV $< m_T <$ 200 GeV to minimize
the systematic uncertainty. From the dependence of the likelihood on 
$\Gamma_W$, we obtain the $W$ boson width and its error as:  
$\Gamma_W = 2.23^{+ 0.15}_{- 0.14}$(stat.) GeV. 
The combined uncertainty, taking the statistical and systematic uncertainties contribution in 
quadrature, yields the result $\Gamma_W = 2.23^{+ 0.15}_{- 0.14}$(stat.)
$\pm~ 0.10$(sys.) GeV$ = 2.23^{+0.18}_{-0.17}$ GeV.
The $\chi^2$ for the best fit is an acceptable 25.9 for 22 degrees of freedom,
corresponding to a probability of 26\%. 
A comparison of the observed spectrum to the 
probability density function in the fitting region through a Kolmogorov-Smirnov test,
which compares the observed cumulative distribution function for a variable
with a specified theoretical distribution, 
yields $\kappa =0.434$, which is evidence of a good fit.

\begin{figure}[!htb]
\epsfxsize = 3.0in  \epsffile{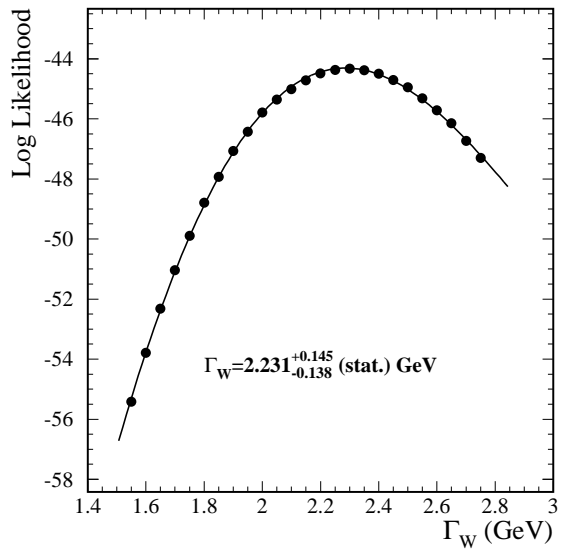}
\caption{ Results of the log-likelihood fit of the data to Monte Carlo
templates for different $\Gamma_W$.}
\label{fig:lik}
\end{figure}

\begin{figure}[!htb]
\begin{center}
\epsfxsize = 3.1in  \epsffile{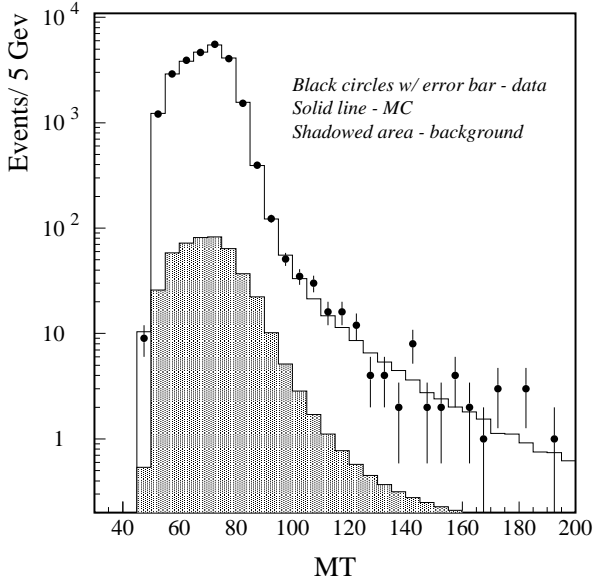}
\caption{ Comparison of data to the Monte Carlo templates for the best fit. 
The black circles with error bars are the data. The solid line of the histogram
corresponds to the MC templates with $\Gamma (W)$ = 2.23~\gev~
normalized to the expected number of $W$ boson events. The shadowed area is the background.}
\label{fig:bestfit}
\end{center}
\end{figure}

Figure~\ref{fig:lik} shows a fit to the likelihood, which
corresponds to a fourth-order polynomial fit that 
determines the peak position. Figure~\ref{fig:bestfit} shows the $m_T$ 
spectrum for the data, the normalized MC sample, and the background.

%

As a consistency check of the fitting method, we also determine 
the $W$ boson width from the ratio of the number of events in the 
fitting region of 90 GeV $\le m_T \le 200$ GeV to the 
number of events in the entire spectrum. 
This yields $\Gamma_W = 2.22 \pm 0.14$(stat.) GeV, compared to
$\Gamma_W = 2.23^{+ 0.15}_{- 0.14}$(stat.) GeV for the independent 
maximum likelihood fit in the same region. All results show good agreement.

Sources of systematic uncertainties in the determination of the $W$ boson width 
are those that can affect the shape of the transverse mass distribution. These 
include the uncertainties from input parameters to the MC program and from 
background estimation. Details can be found in Ref.~\cite{xuthesis}. 
Table~\ref{tab:SYST} lists all the 
sources of systematic uncertainty for the decay width of the $W$ boson.
\newcommand{\error}{$\delta \Gamma_W$(MeV)}
\begin{table}[!htb]
\begin{tabular}{lc}
   Source                   & \error \\
\hline
Hadronic energy resolution  & 55  \\
EM energy scale             & 41  \\
Background ensemble studies & 39  \\
Luminosity slope dependence & 28  \\
EM energy resolution        & 27  \\
PDF                         & 27  \\
Hadronic energy scale       & 22  \\
Background normalization    & 15  \\ 
$W$ boson mass              & 15  \\
Production model            & 12  \\
Radiative correction        & 10  \\
Selection bias              & 10  \\
Angular calibration of $e$ trajectory      & 9    \\
\hline\hline
Total systematic uncertainty & 99 \\
\hline
Total statistical uncertainty & +145 \\
                            & $-138$ \\
\hline\hline
Total uncertainty           & +176 \\
                            & $-170$ \\
\end{tabular}
\caption{Systematic uncertainties and the total uncertainty on the
$W$ boson width measurement.}
\label{tab:SYST}
\end{table}

Comparing to the SM prediction of $\Gamma (W) =2.0921 \pm 0.0025$ GeV,
we find the difference between SM prediction and our meaasurement to
be $0.24 ^{+0.18}_{-0.17}$ GeV, which is the width for the $W$ boson to 
decay into final states other than the two lightest quark doublets and
the three lepton doublets. We set a 95\% confidence level upper limit
on the $W$ boson width to non-SM final states. Assuming the uncertainty
is Gaussian, we set a 95\% confidence level upper limit on the invisible
partial width of the $W$ boson to be 0.59 GeV. Under the assumption that 
there is no correlation between indirect measurement and direct measurement 
of the $W$ boson decay width and within the framework of SM, we can combine 
both analysis and obtain $\Gamma_W = 2.162 \pm 0.062$ GeV. The 95\% 
confidence level upper limit on the invisible partial width of the $W$ 
boson is 0.191 GeV.

\section {Conclusions}

We have directly measured the decay width of the $W$ boson by 
fitting the transverse mass in $W \rightarrow e\nu$ events 
in $p\bar{p}$ collisions at 1.8 TeV, and obtain:

\begin{eqnarray}
\Gamma _W & = & 2.23 ^{+ 0.15}_{- 0.14}(\text{stat.}) \pm 0.10(\text{syst.})~\text{GeV}\\
          & = & 2.23^{+0.18}_{-0.17}~\text{GeV}.
\end{eqnarray}

This result is consistent with the prediction of the standard model.

\section{Acknowledgments}

\input{acknowledgement_paragraph_r1.tex}
\end{document}

%% file: d0_style.tex
\def\pbar{$\overline{p}$}               
\def\ppbar{$p\overline{p} $}            
\def\et{$E_T$}                          
\def\met{\mbox{${\hbox{$E$\kern-0.6em\lower-.1ex\hbox{/}}}_T$}} 
\def\gev{GeV}                           
\def\D0{D\O}                            
\def\etal{{\sl et al.}}                 
\def\d0draft{}
\def\err#1#2#3 {{\it Erratum} {\bf#1},{\ #2} (19#3)}
\def\ib#1#2#3 {{\it ibid.} {\bf#1},{\ #2} (19#3)}
\def\nc#1#2#3 {Nuovo Cim. {\bf#1} ,#2(19#3)}
\def\nim#1#2#3 {Nucl. Instr. Meth. {\bf#1},{\ #2} (19#3)}
\def\np#1#2#3 {Nucl. Phys. {\bf#1},{\ #2} (19#3)}
\def\pl#1#2#3 {Phys. Lett. {\bf#1},{\ #2} (19#3)}
\def\prev#1#2#3 {Phys. Rev. {\bf#1},{\ #2} (19#3)}
\def\prl#1#2#3 {Phys. Rev. Lett. {\bf#1},{\ #2} (19#3)}
\def\rmp#1#2#3 {Rev. Mod. Phys. {\bf#1},{\ #2} (19#3)}
\def\zp#1#2#3 {Zeit. Phys. {\bf#1},{\ #2} (19#3)}

%% file: list_of_authors_r1.tex
%
\author{                                                                      
V.M.~Abazov,$^{23}$                                                           
B.~Abbott,$^{57}$                                                             
A.~Abdesselam,$^{11}$                                                         
M.~Abolins,$^{50}$                                                            
V.~Abramov,$^{26}$                                                            
B.S.~Acharya,$^{17}$                                                          
D.L.~Adams,$^{55}$                                                            
M.~Adams,$^{37}$                                                              
S.N.~Ahmed,$^{21}$                                                            
G.D.~Alexeev,$^{23}$
% ...
\\                                                                            
\vskip 0.30cm                                                                 
\centerline{(D\O\ Collaboration)}                                             
\vskip 0.30cm                                                                 
}                                                                             
\address{                                                                     
\centerline{$^{1}$Universidad de Buenos Aires, Buenos Aires, Argentina}       
\centerline{$^{2}$LAFEX, Centro Brasileiro de Pesquisas F{\'\i}sicas,         
                  Rio de Janeiro, Brazil}                                     
\centerline{$^{3}$Universidade do Estado do Rio de Janeiro,                   
                  Rio de Janeiro, Brazil}                                     
\centerline{$^{4}$Institute of High Energy Physics, Beijing,                  
                  People's Republic of China}                                 
\centerline{$^{5}$Universidad de los Andes, Bogot\'{a}, Colombia}             
\centerline{$^{6}$Charles University, Center for Particle Physics,            
                  Prague, Czech Republic}                                     
\centerline{$^{7}$Institute of Physics, Academy of Sciences, Center           
                  for Particle Physics, Prague, Czech Republic}               
\centerline{$^{8}$Universidad San Francisco de Quito, Quito, Ecuador}         
\centerline{$^{9}$Institut des Sciences Nucl\'eaires, IN2P3-CNRS,             
                  Universite de Grenoble 1, Grenoble, France}                 
\centerline{$^{10}$CPPM, IN2P3-CNRS, Universit\'e de la M\'editerran\'ee,     
                  Marseille, France}                                          
\centerline{$^{11}$Laboratoire de l'Acc\'el\'erateur Lin\'eaire,              
                  IN2P3-CNRS, Orsay, France}                                  
\centerline{$^{12}$LPNHE, Universit\'es Paris VI and VII, IN2P3-CNRS,         
                  Paris, France}                                              
\centerline{$^{13}$DAPNIA/Service de Physique des Particules, CEA, Saclay,    
                  France}                                                     
\centerline{$^{14}$Universit{\"a}t Mainz, Institut f{\"u}r Physik,            
                  Mainz, Germany}                                             
\centerline{$^{15}$Panjab University, Chandigarh, India}                      
\centerline{$^{16}$Delhi University, Delhi, India}                            
\centerline{$^{17}$Tata Institute of Fundamental Research, Mumbai, India}     
\centerline{$^{18}$Seoul National University, Seoul, Korea}                   
\centerline{$^{19}$CINVESTAV, Mexico City, Mexico}                            
\centerline{$^{20}$FOM-Institute NIKHEF and University of                     
                  Amsterdam/NIKHEF, Amsterdam, The Netherlands}               
\centerline{$^{21}$University of Nijmegen/NIKHEF, Nijmegen, The               
                  Netherlands}                                                
\centerline{$^{22}$Institute of Nuclear Physics, Krak\'ow, Poland}            
\centerline{$^{23}$Joint Institute for Nuclear Research, Dubna, Russia}       
\centerline{$^{24}$Institute for Theoretical and Experimental Physics,        
                   Moscow, Russia}                                            
\centerline{$^{25}$Moscow State University, Moscow, Russia}                   
\centerline{$^{26}$Institute for High Energy Physics, Protvino, Russia}       
\centerline{$^{27}$Lancaster University, Lancaster, United Kingdom}           
\centerline{$^{28}$Imperial College, London, United Kingdom}                  
\centerline{$^{29}$University of Arizona, Tucson, Arizona 85721}              
\centerline{$^{30}$Lawrence Berkeley National Laboratory and University of    
                  California, Berkeley, California 94720}                     
\centerline{$^{31}$University of California, Davis, California 95616}         
\centerline{$^{32}$California State University, Fresno, California 93740}     
\centerline{$^{33}$University of California, Irvine, California 92697}        
\centerline{$^{34}$University of California, Riverside, California 92521}     
\centerline{$^{35}$Florida State University, Tallahassee, Florida 32306}      
\centerline{$^{36}$Fermi National Accelerator Laboratory, Batavia,            
                   Illinois 60510}                                            
\centerline{$^{37}$University of Illinois at Chicago, Chicago,                
                   Illinois 60607}                                            
\centerline{$^{38}$Northern Illinois University, DeKalb, Illinois 60115}      
\centerline{$^{39}$Northwestern University, Evanston, Illinois 60208}         
\centerline{$^{40}$Indiana University, Bloomington, Indiana 47405}            
\centerline{$^{41}$University of Notre Dame, Notre Dame, Indiana 46556}       
\centerline{$^{42}$Iowa State University, Ames, Iowa 50011}                   
\centerline{$^{43}$University of Kansas, Lawrence, Kansas 66045}              
\centerline{$^{44}$Kansas State University, Manhattan, Kansas 66506}          
\centerline{$^{45}$Louisiana Tech University, Ruston, Louisiana 71272}        
\centerline{$^{46}$University of Maryland, College Park, Maryland 20742}      
\centerline{$^{47}$Boston University, Boston, Massachusetts 02215}            
\centerline{$^{48}$Northeastern University, Boston, Massachusetts 02115}      
\centerline{$^{49}$University of Michigan, Ann Arbor, Michigan 48109}         
\centerline{$^{50}$Michigan State University, East Lansing, Michigan 48824}   
\centerline{$^{51}$University of Nebraska, Lincoln, Nebraska 68588}           
\centerline{$^{52}$Columbia University, New York, New York 10027}             
\centerline{$^{53}$University of Rochester, Rochester, New York 14627}        
\centerline{$^{54}$State University of New York, Stony Brook,                 
                   New York 11794}                                            
\centerline{$^{55}$Brookhaven National Laboratory, Upton, New York 11973}     
\centerline{$^{56}$Langston University, Langston, Oklahoma 73050}             
\centerline{$^{57}$University of Oklahoma, Norman, Oklahoma 73019}            
\centerline{$^{58}$Brown University, Providence, Rhode Island 02912}          
\centerline{$^{59}$University of Texas, Arlington, Texas 76019}               
\centerline{$^{60}$Texas A\&M University, College Station, Texas 77843}       
\centerline{$^{61}$Rice University, Houston, Texas 77005}                     
\centerline{$^{62}$University of Virginia, Charlottesville, Virginia 22901}   
\centerline{$^{63}$University of Washington, Seattle, Washington 98195}       
}                                                                             

%% file: acknowledgement_paragraph_r1.tex
%
We thank the staffs at Fermilab and collaborating institutions, 
and acknowledge support from the 
Department of Energy and National Science Foundation (USA),  
Commissariat  \` a L'Energie Atomique and 
CNRS/Institut National de Physique Nucl\'eaire et 
de Physique des Particules (France), 
Ministry for Science and Technology and Ministry for Atomic 
   Energy (Russia),
CAPES and CNPq (Brazil),
Departments of Atomic Energy and Science and Education (India),
Colciencias (Colombia),
CONACyT (Mexico),
Ministry of Education and KOSEF (Korea),
CONICET and UBACyT (Argentina),
The Foundation for Fundamental Research on Matter (The Netherlands),
PPARC (United Kingdom),
Ministry of Education (Czech Republic),
A.P.~Sloan Foundation,
NATO, and the Research Corporation.

%% file: prd_v50.bbl
\begin{references}
%
\bibitem[*]{lehner}
Also at University of Zurich, Zurich, Switzerland.
\bibitem[\dag]{przybycien}
Also at Institute of Nuclear Physics, Krakow, Poland.
%
\vskip 0.25cm


\bibitem{SM_Glashow}
   S. Glashow, Nucl. Phys. {\bf22}, 579 (1961).

\bibitem{SM_Weinberg}
   S. Weinberg, Phys. Rev. Lett. {\bf19}, 1264 (1967) .

\bibitem{SM_Salam}
   A. Salam in {\it Elementary Particle Theory}, edited by N. Svartholm (Almquist 
   and Wiksells, Stockholm, 1969), 367.

\bibitem{QCD_Bardeen}
   W. Bardeen, H. Fritzsch, M. Gell-Mann in {\it Scale and Conformal Symmetry in 
   Hadron Physics}, edited by R. Gatto (Wiley, New York, 1973), 139.

\bibitem{QCD_Gross}
   D. Gross and F. Wilczek, Phys. Rev. D {\bf8}, 3633 (1973).

\bibitem{QCD_Weinberg}
   S. Weinberg, Phys. Rev. Lett. {\bf31}, 494 (1973).


\bibitem{UA1_W_discovery}
   UA1 Collaboration, G.~Arnison \etal, Phys. Lett. B {\bf122}, 103 (1983).

\bibitem{UA2_W_discovery}
   UA2 Collaboration, P.~Bagnaia \etal, Phys. Lett. B {\bf122}, 476 (1983).

\bibitem{UA1_Z_discovery}
   UA1 Collaboration, G.~Arnison \etal, Phys. Lett. B {\bf126}, 398 (1983).

\bibitem{UA2_Z_discovery}
   UA2 Collaboration, P.~Bagnaia \etal, Phys. Lett. B {\bf129}, 130 (1983).


\bibitem{epj:groom}
   D. E. Groom \etal, Eur. Phys. J. C {\bf15}, 1-878 (2000).

\bibitem{prd:rosner}
   J. Rosner, M. Worah, and T. Takeuchi, Phys. Rev. D {\bf49}, 1363 (1994).

\bibitem{fconst}
   R. M. Barnett \etal, Phys. Rev. D {\bf54}, 1 (1996).

\bibitem{cdf:mu}
   CDF Collaboration, F. Abe \etal, Phys. Rev. D{\bf43}, 2070 (1991).

\bibitem{wmassprd}
   \Dzero\ Collaboration, B.~Abbott \etal, Phys. Rev. D {\bf 58}, 092003 (1998). 

\bibitem{wmassl3}
   L3 Collaboration, M.~Acciarri \etal, Phys. Lett. B {\bf 413}, 176 (1997). 

\bibitem{wmassalph}
   ALEPH Collaboration, R.~Barate \etal, Phys. lett. B {\bf 422}, 384 (1998). 

\bibitem{wmassopal}
   OPAL Collaboration, K.~Ackerstaff \etal, Eur. Phys. J. C {\bf 1}, 395 (1998). 

\bibitem{wmassdlph}
   Delphi Collaboration, P.~Abreu \etal, Eur. Phys. J. C {\bf 2}, 581 (1998). 

\bibitem{mrsd:plb} 
   A. D. Martin, R. G. Roberts and W. J. Stirling, Phys. Lett. B {\bf306}, 
   147 (1993) and {\bf309}, 492 (1993).

\bibitem{cern:lep} 
   LEP Electroweak Working Group, CERN Report No. CERN-EP-2001-098, 
   hep-ex/0112021.

\bibitem{pl:ua1}
   UA1 Collaboration, C. Albajar \etal, Phys. Lett. B {\bf253}, 503 (1991).

\bibitem{pl:ua2} 
   UA2 Collaboration, J. Alitti \etal, Phys. Lett. B {\bf276}, 365 (1992).

\bibitem{cdf:ratio} 
   CDF Collaboration, F. Abe  \etal, Phys. Rev. D {\bf52}, 2624 (1995).

\bibitem{wzcsprd} 
   \D0 Collaboration, B. Abbott \etal, Phys. Rev. D {\bf61}, 072001 (2000).

\bibitem{vdb:rjnp}
   V. D. Barger and R. J. N. Phillips, in $Collider Physics$, Volume 71 of Frontiers 
   in Physics (Addison-Wesley, 1987).

\bibitem{cdf:wwidth}
   CDF Collaboration, F. Abe  \etal, Phys. Rev. Lett. {\bf85}, 3347 (2000). 
   CDF measured the $W$ boson width in both the $e\nu$ and $\mu\nu$ channels. 
   The number reported is their combined result.

\bibitem{xuthesis} 
   Qichun Xu, 
   Ph. D. Thesis, the University of Michigan, 2001 (unpublished).
   $http://www\-d0.fnal.gov/results/publications\_talks/thesis/$ $xu/qichun\_thesis.html$. 

\bibitem{etadef} 
   The origin of the coordinate system is the reconstructed position of $p\bar{p}$ 
   interaction when describing the interaction, and the geometrical center of the 
   detector when describing the detector. It refers to the detector here.

\bibitem{d0nim}
    \Dzero\ Collaboration, S.~Abachi {\it et al.},
    Nucl. Instr. and Methods in  Phys. Res. A {\bf 338}, 185 (1994).

\bibitem{topmassprd}
   \Dzero\ Collaboration, B.~Abbott \etal, Phys. Rev. D {\bf 58}, 052001 (1998). 

\bibitem{miguel} 
   \Dzero\ Collaboration, V. M. Abazov \etal, Phys. Lett. B 513, 292 (2001).

\bibitem{covprd}
   \Dzero\ Collaboration, S.~Abachi \etal, Phys. Rev. D {\bf 52}, 4877 (1995).


\bibitem{flattum} 
   Eric M. Flattum, 
   Ph. D. Thesis, Michigan State University, 1996 (unpublished),
   $http://www\-d0.fnal.gov/results/publications\_talks/thesis/$ $flattum/eric\_thesis.html$. 

\bibitem{wmass:thesis}
   Ian Malcolm Adam, Ph. D. Thesis, Columbia University, 1997 (unpublished), 
   $http://www\-d0.fnal.gov/results/publications\_talks/thesis/$ $adam/ian\_thesis\_all.html$. 

\bibitem{resb} 
   C. Balazs and C. P. Yuan, Phys. Rev. D {\bf 56}, 5558 (1997).

\bibitem{ly} 
   G. A. Ladinsky and C. P. Yuan, Phys. Rev. D {\bf 50}, 4239 (1994).

\bibitem{ARtheory}
   P. B. Arnold and M.H. Reno, Nucl. Phys. {\bf B319}, 37 (1989);
   {\bf B330}, 284E (1990); R. J.~Gonsalves, J. Pawlowski, and C-F. Wai, 
   Phys. Rev. D {\bf 40}, 2245 (1989).

\bibitem{parton}
   \Dzero\ Collaboration, S. Abachi \etal, Phys. Rev. Lett. {\bf77}, 3309 (1996); 
   \Dzero\ Collaboration, B. Abbott \etal, Phys. Rev. D {\bf 58}, 012002 (1998).

\bibitem{MRSA}
   A. D. Martin, W. J. Stirling, R. G. Roberts, Phys. Lett. B {\bf354}, 155 (1995).

\bibitem{CTEQ}
   The Coordinated Theoretical-Experimental Project on QCD, 
   $http://www.phys.psu.edu/cteq$.

\bibitem{MRST}
   A. D. Martin, R. G. Roberts and W. J. Stirling, R. R. Thorne, hep-ph/0110215.

\bibitem{bkgfitpara}
   We found the fitting parameters as $a_0 = (3.9153 \pm 0.0012) \times 10^{1}$,
   $a_1 = (-7.5100 \pm 0.0044) \times 10^{-1}$, $a_2 = (4.7087 \pm 0.0041) \times 10^{-3}$ and 
   $a_3 = (-1.00461 \pm 0.00095) \times 10^{-5}$.

\bibitem{herwig}
   G.~Marchesini \etal, Comput. Phys. Commun. {\bf 67}, 465 (1992).

\bibitem{geant}
   F.~Carminati \etal, {\it \GEAN\ Users Guide},
   CERN Program Library W5013, 1991 (unpublished).

\end{references}
